**A Comparison between two methods for the measurement of the Planck constant $h$: the watt balance and the superconductor electromechanical oscillator methods.**

Osvaldo F Schilling

Departamento de Física, UFSC, 88040-900, Florianópolis, SC. Brazil.

email: osvaldof@mbox1.ufsc.br

phone: +55 48 37216832 ext 4101

fax: +55 48 37219946

cell phone: +55 48 99696601

skype username: ofschilling

**Introduction**

The purpose of this note is to make a brief analysis of the physical principles upon which two methods for relating the mass of an object to fundamental physical constants are based. The two methods are, namely, the watt balance method (WB) [1,2], and a still untested experimental technique based upon the "Superconductor Electromechanical Oscillator" ( SEO)[3,4]. We show that both these methods are governed by similar equations.

**The Watt balance method**

This analysis is based upon the detailed discussion of the WB method available in the BIPM site [1] and upon the report of Steiner et al. [2] describing the recent efforts in NIST for the measurement of $h$ with their watt balance. According to [1,2], the WB method requires two complementary experiments ( Modes I and II, as described in [2]) to produce a connection between a standard mass and $h$.

Mode I: A horizontal coil moves in the presence of a magnetic field gradient produced by a superconducting solenoid. The induced voltage in the coil $U$ is measured against the Josephson standard. The velocity of the coil $v$ is obtained by measuring its position against time. The equation for $U$ is derived from Faraday's induction law:

$$U = -v \, d\Phi/dx \qquad (1)$$

Mode II: The coil transports a current $I$ fed by a current source, and the magnetic force produced upon this current by the magnetic field gradient of the solenoid balances a standard weight $F = mg$. The equation describing such balance is:

$$F = -I \, d\Phi/dx \tag{2}$$

Provided the flux gradients can be eliminated in (1) and (2), one obtains the equality between the measured mechanical power $Fv$ and electrical power $UI$, with deviations $\varepsilon$ attributed to corrections to be made in the units adopted.

$$(Fv)_{SI}/(UI)_{90} = (mgv)_{SI}/(UI)_{90} = 1 + \varepsilon \tag{3}$$

which can be rewritten in the form[1,2]:

$$m = h(\, K_{J-90}^2 \, R_{K-90}/4)(U_{90} \, I_{90})/(gv) \tag{4}$$

The subscripts SI and 90 refer respectively to SI units adopted in mechanical power measurements, and to electrical standards adopted in 1990. $R_{K-90}$ is the von Klitzing constant and $K_{J-90}$ is the Josephson constant. That is, from the knowledge of $m$ obtained by a comparison with the Pt-Ir artifact in Sevres, a value for the Planck $h$ is obtained through the other measurements.

**The Superconductor Electromechanical Oscillator Method (SEO)[3,4].**

The SEO has been described in great detail in [3,4]. We will give special attention to the comparison between the physical principles and show that they are the same for both methods. Let´s consider the model experimental setup described in Figure 1 of [4]. A rectangular type-II superconductor loop of mass $m$ is subjected to magnetic fields $B_1$ and $B_2$ from a system of magnets. In the Figure a smaller magnet concentrates the field $B_2$ upon part of the lower horizontal leg of the loop. We will assume $B_1 > B_2 > B_{c1}$ ($B_{c1}$ is the superconductor lower critical field). Here $a$ is the size of the region in which $B = B_2$ in Fig. 1. The loop will move with speed $v$ described by Newton´s Law:

$$m \, dv/dt = mg - iaB_0 \tag{5}$$

We introduce the parameter $B_0 \equiv B_1 - B_2$ to simplify the notation. Note the analogy between (5) and (2), although in this case the equilibrium is dynamic rather than static. The displacement of a <u>normal conducting</u> loop gives rise to an induced electromotive force $U$, given by Faraday's Induction Law:

$$U = -d\Phi_m/dt - L\, di/dt \qquad (6)$$

Here $\Phi_m$ is the magnetic flux from the magnets that penetrates the rectangular area bound by the loop, and $L$ is the self-inductance of the loop. $\Phi_m + Li$ is the total magnetic flux within the loop area, which is <u>conserved since the loop is superconducting</u> and we neglect for the moment any other dissipative processes. Therefore, the electromotive force $U$ in (6) will be considered <u>zero</u>. Equation (6) plays for the SEO the same role that (1) does for the WB method. From Figure 1, $d\Phi_m/dt = -B_0 a v$, and thus from (6) one obtains a relation between $v$ and $di/dt$. Taking the time derivative of (5) and eliminating $di/dt$ from (6) one obtains:

$$d^2v/dt^2 = -(B_0^2 a^2/Lm)\, v \qquad (7)$$

which is the differential equation obeyed by the velocity of a harmonic oscillator. This means that the loop should perform an *ideal* oscillating motion under the action of the external and magnetic forces in the absence of losses. Assuming zero initial speed and an initial acceleration $g$, eq. (7) can be solved:

$$v(t) = (g/\Omega)\, \sin(\Omega t) \qquad (8)$$

From (7), the natural frequency of the oscillations is $\Omega = B_0 a/(mL)^{1/2}$. It is possible then to combine (5)–(6) to obtain an equation for the current $i(t)$:

$$(B_0 a/\Omega^2)\, d^2i/dt^2 = mg - B_0 a i \qquad (9)$$

whose solution is

$$i(t) = (mg/(B_0 a))(1 - \cos(\Omega t)) \qquad (10)$$

for $i(0) = di/dt(0) = 0$. From eq. (10) we conclude that the supercurrent $i$ induced in the coil never changes sign, and it looks like a rectified current. The amplitude of the ac component of the current is the same as the dc average term, and we define it as $i_0 \equiv mg/(B_0 a)$.

As the loop is released from rest, the assumed perfect flux and energy conservations will make this initial position the uppermost point of its trajectory (measured from the middle point of its oscillating vertical trajectory and negative upwards), with $x(0) = -x_0 = -g/\Omega^2$. The loop position is described by the equation

$$x(t) = -(g/\Omega^2)\cos(\Omega t) \tag{11}$$

The amplitude of the oscillating motion is $x_0 = g/\Omega^2$, which may be quite small since it is inversely proportional to $\Omega^2$.

The initial gravitational potential energy must match the sum of the gravitational, magnetic and kinetic energies at any time, resulting in an energy conservation equation:

$$0 = mg(x(0) - x) + \tfrac{1}{2} mv^2 + \tfrac{1}{2} Li^2 \tag{12}$$

Here the first two terms on the right side are the mechanical energy terms and the last one the magnetic energy. In order to obtain an expression comparable to (3) for the WB method we take the time derivatives of (12) for the powers $P$. It comes at once that $P_{mec} = -P_{mag}$, so that

$$mgx_0\Omega = L i_0^2 \Omega = (L i_0 \Omega) i_0 \tag{13}$$

Since $x_0 \Omega = v_0$, the maximum speed of oscillations, and $(L i_0 \Omega)$ has the dimensions of voltage (say, $V$), we ended up with an equation completely analogous to (3) for the WB method.

$$mgv_0/(V i_0) = 1 \tag{14}$$

The corrections in (14) will depend on the particular techniques and standards to be adopted on measuring the parameters in (13)

**Analysis**

The measured parameters needed in (14) are the gravity acceleration, the amplitude of the ac current (or dc component) $i_0$, and its frequency $\Omega$, and the loop $L$. It has been pointed out by Franco Cabiati that there is an advantage of adopting an oscillatory motion, performing both the "weighing experiment" and the "moving experiment" of the watt balance simultaneously in a single experiment. Furthermore, the frequency of the oscillation can replace one of the two electrical parameters (voltage or current) that link the Planck constant through the Josephson and von Klitzing effects. In the case of the SEO system, it must be assumed that the oscillation is sufficintly slow and large to allow for accurate measurement of kinematic parameters on which mechanical power depends.

However, (14) is entirely based on the hypothesis of complete flux and energy conservation, which is unattainable in practice. Ref.[4] was fully dedicated to such discussion. In that paper we took the example of a 5x5 sq. cm Nb-48% Ti loop made of a wire 0.6 mm thick ( this is an extremely hard alloy and the wire will not flex). To understand all the argumentation in [3,4] a considerable expertise on the technical properties of "hard" superconductors is necessary. In particular, it must be understood that the system will only work because it should be possible to make the currents ( on the order of 1 A) flow in a micrometer-thin layer close to the surface of the wire, with no hysteresis losses.  Tiny losses associated with the oscillations of normal electrons in the cores of the magnetic flux-lines (FL) that penetrate the wires still remain. However, the most important dissipative effect ends up being the drag of the loop against the atmosphere in the cryostat[4]. We made the calculations for T= 0.05K and pressure of $10^{-8}$ Torr ( high-vacuum conditions), with $B_0$= 0.3 Tesla, resulting in $x_0$= 8 µm and $f$ = $\Omega/2\pi$ = 178 Hz. For these conditions the two loss mechanisms give similar results and we conclude that the quality factor $Q$ attainable would reach $2\times10^{10}$. For real applications it might be possible to work at 4K, with worse vacuum conditions, and accept inhomogeneities in the magnetic field imposed to the loop. This would take $Q$ to about $10^9$. Therefore, 1 part in $10^9$ should be the degree of stability expected in the frequency $\Omega$ that appears in (14). We note that (14) also dependends upon the self-inductance $L$. This is a parameter that should in this case be measured at the work temperature in an independent procedure. The current in the loop might be obtained through a superconducting current comparator wound around one of the vertical legs of the loop, as pointed out by John Gallop of the British NPL.

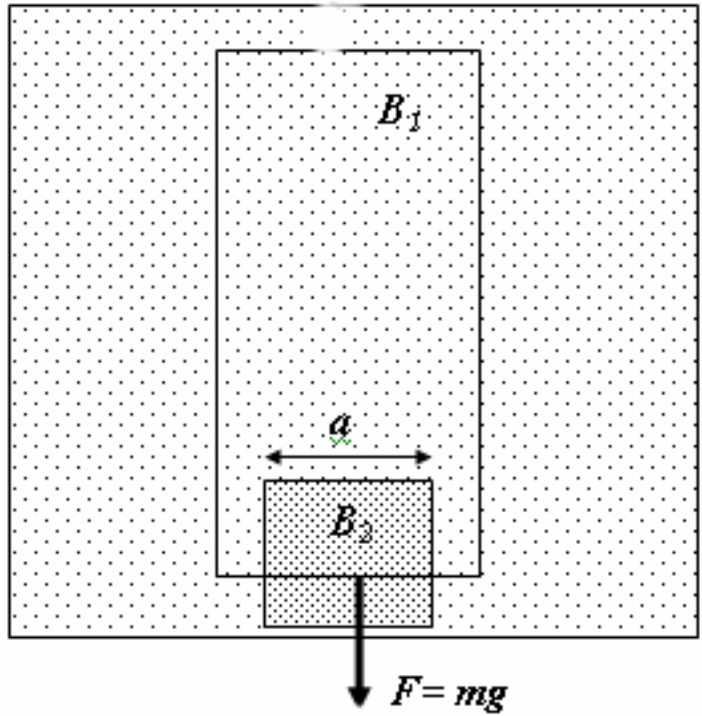

Figure 1